\documentclass[prd,twocolumn]{revtex4}\begin{document}\preprint{}
\title{Stochastic inflaton wave equation from an expanding environment }
\author{ Z. Haba\\
Institute of Theoretical Physics, University of Wroclaw,\\ 50-204
Wroclaw, Plac Maxa Borna 9, Poland}
\email{zbigniew.haba@uwr.edu.pl}
\begin{abstract}We discuss the inflaton  $\phi$ in an environment  of
scalar fields $\chi_{n}$ on  flat and curved manifolds. We average
over the environmental  fields $\chi_{n}$. We study a contribution
of superhorizon $k<<aH$ as well as subhorizon  $k>>aH$ modes
$\chi_{n}({\bf k})$. As a result we obtain a stochastic wave
equation with a friction and noise. We show that in the subhorizon
regime in field theory a finite number of fields is sufficient to
produce a friction and diffusion owing to the infinite number of
degrees of freedom corresponding to different ${\bf k}$ in
$\chi_{n}({\bf k})$. We investigate the slow roll  and the
Markovian approximations to the stochastic wave equation. A
determination of the metric from the stochastic
Einstein-Klein-Gordon  equations is briefly discussed.
\end{abstract}
\maketitle

\section{Introduction}
Inflation is generated by an interaction of gravity with a scalar
field (inflaton) \cite{starfirst}\cite{guth}
\cite{linde}\cite{mukhanov}\cite{sasaki}  . In a purely
Hamiltonian system it is  impossible to reach a thermodynamic
equilibrium exhibited in the CMB radiation. The equilibration can
be achieved by an interaction with an environment of other fields
(heat bath). Then, the temperature of the inflaton subsystem is
decreasing during inflation.  In the standard (cold inflation)
approach  a mechanism of reheating
\cite{reheating}\cite{reheating2} is applied to raise the
temperature. The presence of the environment limits the decrease
of temperature
\cite{fang}\cite{warm}\cite{bererainteraction}.Berera
\cite{berera} described the effect of the environment  by a
stochastic  modification of the inflaton equation. He derived a
stochastic equation from a linear interaction with an infinite set
of scalar fields in a heat bath. The model follows the well-known
derivation of the Brownian motion from dynamical systems
\cite{ford}\cite{kac}. We have obtained the same stochastic
equation in the low momentum regime ($k\simeq 0$) by different
mathematical methods in \cite{adv}. In \cite{berera}\cite{adv} the
expansion of the environment has been
 neglected in the derivation (as we set $a^{-2}k^{2}\simeq 0$). What was essential for
the result was  an infinite set of fields with masses proportional
to couplings. Such a relation allows to apply the Markov
approximation. In Appendix B of ref.\cite{berera} Berera derives
the same diffusion equation if there is only one field in the
environment but $k>>aH$.

In this paper we investigate models of an interaction with the
environment in more detail starting from models in the Minkowski
space. We consider some generalizations of the models of
\cite{berera}\cite{adv}. The main new result concerns the
calculation of  the effect of the subhorizon  modes $k>>aH$ on the
wave equation of the inflaton. We show that if the momentum and
scale-dependent term in the scalar-field equations is dominating
(in the subhorizon regime $k>>aH$ ) then we obtain a different
diffusive inflaton equation than the one resulting from averaging
over superhorizon modes (in disagreement with \cite{berera}). The
appearance of the diffusive behaviour in this model is similar to
the one discussed by Starobinsky  and Vilenkin
\cite{star}\cite{vilenkin}\cite{vilenkin2}\cite{vilenkin3} when
modes with large $k$ of the quantum scalar field lead to a
diffusive behaviour of the remaining superhorizon modes $k<<aH$ in
an expanding universe. In such a case those authors were able to
derive a stochastic wave equation describing quantum fluctuations
in an arbitrary inflaton potential. Quantum as well as thermal
fluctuations determine the CMB spectrum which can be compared with
observations \cite{a1}\cite{a2}.

 The plan of this paper is the
following. In sec.2 we review a minor generalization of the model
of refs.\cite{berera}\cite{adv}. In sec.3 we discuss in detail the
model in Minkowski space-time.In sec.4 we study the environmental
fields for subhorizon momenta in de Sitter space. In sec.5 we
explore an expansion in a fixed homogeneous metric which is close
to exponential. We eliminate the environmental fields, average
over the initial values and obtain a stochastic wave equation for
the inflaton with a friction and noise. We briefly discuss the
slow roll conditions resulting from our modification of the
stochastic equation for warm inflation. We show that in some
approximations the resulting wave equation for the inflaton can be
treated as a Markovian stochastic wave equation. In sec.6 we
briefly discuss the Friedman equation which determines the
homogeneous metric ( the Hubble variable as a function of the
inflaton field) .
 We believe that the averaging over the initial values simulates some quantum
effects in cosmological models as the quantum field theory at
finite temperature tends to the classical field theory with random
initial conditions distributed according to the classical Gibbs
law. The thermal noise in the inflaton equation can play a similar
role as the scalar quantum fluctuations  (discussed  in
\cite{mukhanov}) which together with the quantum fluctuations of
the gravitational field during de Sitter expansion (calculated
earlier  in \cite{starex}) lead to the correct evaluation of the
power spectrum  (this can be done by means of the methods
developed in \cite{mar} \cite{starquant}) .
\section{Scalar fields interacting linearly with an environment}
We consider the Lagrangian which is an extension of the well-known
oscillator model discussed in \cite{ford}\cite{kac}
\begin{equation}\begin{array}{l}
{\cal L}=\frac{1}{2}\partial_{\mu}\phi\partial^{\mu}\phi
-V(\phi)\cr+\sum_{n}(\frac{1}{2}\partial_{\mu}\chi_{n}\partial^{\mu}\chi_{n}
-\frac{1}{2}m_{n}^{2}\chi_{n}\chi_{n}-\lambda_{n}U(\phi)\chi_{n}),
\end{array}\end{equation}where $U(\phi)$ is a certain interaction. $U(\phi)=\phi$ is a simple choice
(considered in \cite{berera}\cite{adv}) but the results of
averaging over $\chi_{n}$ do not depend essentially on $U$. The
number of particles in the classical mechanics of
\cite{ford}\cite{kac} is infinite. Only in the limit of an
infinite number of degrees of freedom the irreversible diffusive
behaviour can appear. In field theory even with a finite number of
fields we have an infinite number of degrees of freedom (the
spatial Fourier modes). Averaging over those modes can lead to a
diffusive dynamics. Equations of motion read
\begin{equation}
g^{-\frac{1}{2}}\partial_{\mu}(g^{\frac{1}{2}}\partial^{\mu}\phi)=-V^{\prime}-U^{\prime}(\phi)\sum_{n}\lambda_{n}\chi_{n},
\end{equation}
\begin{equation}
g^{-\frac{1}{2}}\partial_{\mu}(g^{\frac{1}{2}}\partial^{\mu}\chi_{n})+m_{n}^{2}\chi_{n}\equiv
{\cal M}_{n}\chi_{n}=-\lambda_{n}U(\phi),
\end{equation}where $g_{\mu\nu}$ is the metric tensor and $g=\vert \det[g_{\mu\nu}]\vert$.
We can consider classical as well as quantum systems (2)-(3).
Eliminating the quantized fields $\chi$ we obtain a quantum
version of the environmental noise. The quantum noise can
approximate the quantum fluctuations of scalar and gravitational
fields as it does in the e-fold time \cite{mar} \cite{starquant}.

In the flat expanding metric
\begin{equation}
ds^{2}=dt^{2}-a^{2}d{\bf x}^{2}
\end{equation}
 eq.(3) reads
\begin{equation}
\partial_{t}^{2}\chi_{n}+3H\partial_{t}\chi_{n}-a^{-2}\triangle\chi_{n}+m_{n}^{2}\chi_{n}=-\lambda_{n}U(\phi).
\end{equation} where $H=a^{-1}\partial_{t}a$.
We can solve eq.(3) for $\chi_{n}$
\begin{equation}
\chi_{n}={\cal A}_{n}\chi_{n}^{cl}-\lambda_{n}\int dx^{\prime}
G_{n}(x,x^{\prime})U(x^{\prime}),\end{equation} where we denote
$U(x)=U(\phi(x))$, $G_{n}$ is the Green function of the operator
${\cal M}_{n}$, ${\cal A}_{n}$ is any operator commuting with
${\cal M}_{n}$ and $\chi_{n}^{cl}$ are  solutions of the
homogeneous equation ${\cal M}_{n}\chi^{cl}_{n}=0$. When we insert
$\chi_{n}$ of eq.(6) in eq.(2) then it takes the form
\begin{equation}\begin{array}{l}
g^{-\frac{1}{2}}\partial_{\mu}(g^{\frac{1}{2}}\partial^{\mu}\phi)+V^{\prime}=U^{\prime}\sum_{n}\lambda_{n}^{2}G_{n}
U  +U^{\prime}\eta\equiv \delta\phi +U^{\prime}\eta,
\end{array}\end{equation} where
\begin{equation}
\eta=-\sum_{n}{\cal A}_{n}\lambda_{n}\chi^{cl}_{n}.
\end{equation}In the homogeneous metric (4) we can take
the spatial Fourier transform of eq.(3). Then, the operators
${\cal A}_{n}$ become just functions $A_{n}({\bf k})$. In
classical field theory (with the energy-momentum tensor
$T^{\mu\nu}$ ) in  Minkowski space $\int T^{00}(\chi)d{\bf x}$ is
a constant of motion as long as there is no coupling to $\phi$.
Then, the Gibbs distribution at temperature $\beta^{-1}$ is
$d\chi\exp(-\beta\int T^{00}(\chi)d{\bf x})$. A natural covariant
generalization to a manifold is the weight factor for classical
configurations of the form\begin{equation}
\exp(-\int^{\beta}d^{4}x\sqrt{g}T^{00}(\chi))
\end{equation}
where the integral $d^{4}x$ is over a volume $\beta$  in
space-time. There is an analog in quantum statistical mechanics to
the formula (9) which is applied as a statistical operator in
quantum non-equilibrium statistical mechanics. For a homogeneous
metric (4) the volume $\beta$ in eq.(9)  can be expressed as a
time interval $[0,\beta]$ times the space volume. After averaging
over the solutions $A_{n}\chi^{cl}_{n}(t,{\bf k})$ the correlation
function of the noise is
\begin{equation}\begin{array}{l}
\langle \eta(x)\eta(y)\rangle=\int d{\bf k}d{\bf
k}^{\prime}\langle A_{m}^{*}({\bf k})A_{n}({\bf
k}^{\prime})\chi^{cl}_{n}(k^{\prime},t^{\prime})\chi^{cl*}_{m}(k,t)\rangle
\cr \lambda_{n}\lambda_{m}\exp(i{\bf k}{\bf x}-i{\bf
k}^{\prime}{\bf y}).\end{array}
\end{equation}
 In this way we
obtain a stochastic wave equation (7)with a friction $\delta\phi$
and the noise $\eta$. There is some arbitrariness in the choice of
the potentials $U$ and $V$, the number of fields $\chi$, the
masses and the couplings. In \cite{berera}\cite{adv}
$U(\phi)=\phi$, an infinite set of fields is chosen and the masses
are proportional to the couplings. As shown in \cite{adv} if we
ignore the $k$ dependence of ${\cal M}_{n}$ (superhorizon domain)
then we obtain a wave equation of warm inflation
\cite{fang}\cite{warm} with the friction proportional to
$\partial_{t}\phi$ and the noise $\eta$ as the white noise. In the
Appendix B of ref.\cite{berera} one field $\chi$ with large
momenta is discussed. It is claimed that its effect is the same as
an infinite number of $\chi$ fields leading to an alternative
derivation of the same stochastic equation for inflaton. We
disagree with this claim. For low momenta of the environmental
fields  another stochastic equation (studied in \cite{invmeasure})
appears than for large momenta. We study these equations in detail
in this paper begining with the simple case of the Minkowski space
in the next section.
\section{Environment in the Minkowski background}
In Minkowski space $a=1$  .  We can solve eq.(3)  exactly for
Fourier transforms
\begin{equation}\begin{array}{l}
\chi_{n}(t,{\bf k})=\cos(\omega_{n}t)\chi_{n0}({\bf k}) +
\sin(\omega_{n}t)\omega_{n}^{-1}\Pi_{n}({\bf k}) \cr
-\lambda_{n}\int_{0}^{t}\sin(\omega_{n}(t-s))\omega_{n}^{-1}U(s,{\bf
k})ds\cr\equiv
\omega_{n}^{-\frac{1}{2}}(A_{n}\exp(i\omega_{n}t)+A_{n}^{*}\exp(-i\omega_{n}t)
-\lambda_{n}G_{n}U\end{array}
\end{equation}
where $(\chi_{n0}({\bf k}),\Pi_{n}({\bf k}))$ are the initial
values for the wave equation, $U(s,{\bf k})$ means the Fourier
transform of $U(s,{\bf x})$ and
\begin{equation} \omega_{n}^{2}=m_{n}^{2}+{\bf k}^{2}.
\end{equation}
Inserting (11) in eq.(2) we obtain an explicit formula
\begin{equation}\begin{array}{l}
\partial_{\mu}\partial^{\mu}\phi+V^{\prime}=\cr\sum_{n}\lambda_{n}^{2}
U^{\prime}\int d{\bf k}\exp(i{\bf k}{\bf
x})\int_{0}^{t}\sin(\omega_{n}(t-s))\omega_{n}^{-1}U(s,{\bf
k})ds\cr+U^{\prime}\eta,\end{array}
\end{equation}
where
\begin{equation}\begin{array}{l}\eta=-\sum_{n}
\lambda_{n}(\cos(\omega_{n}t)\chi_{n0}({\bf k}) +
\sin(\omega_{n}t)\omega_{n}^{-1}\Pi_{n}({\bf k})) \cr\equiv
-\sum_{n}\lambda_{n}\omega_{n}^{-\frac{1}{2}}(A_{n}\exp(i\omega_{n}t)+A_{n}^{*}\exp(-i\omega_{n}t)).
\end{array}\end{equation}
The friction term could also be expressed in a Lorentz covariant
form (eqs. (2)-(3) are Lorentz invariant in Minkowski space, hence
solutions should preserve this invariance)
\begin{displaymath}\begin{array}{l}
-i\sum_{n}\lambda_{n}^{2}U^{\prime}\int_{0}^{t} \cr\int
dk\exp(-ik_{\mu}(x^{\mu}-y^{\mu}))sign(k)\delta(k^{2}-m_{n}^{2})U(s,y)dy_{0}d{\bf
y}\end{array}\end{displaymath} where $sign(\tau)$ is an odd
function with $sign (\tau) =1$ for $\tau>0$ and $x=(t,{\bf x})$.

The Hamiltonian for the wave fields $\chi_{n}$ is
($\Pi=\partial_{\tau}\chi$)
\begin{equation}
{\cal H}_{n}=\frac{1}{2}\int d{\bf x}
(\Pi^{2}+(\nabla\chi)^{2}+m_{n}^{2}\chi^{2}) =\int d{\bf k}
A_{n}^{*}A_{n}\omega_{n}\end{equation} Then, the expectation value
with respect to the Gibbs density (9) is
\begin{displaymath}
\langle A^{*}_{n}({\bf k})A_{m}({\bf
k}^{\prime})\rangle=\beta^{-1}\delta_{mn}\delta({\bf k}-{\bf
k}^{\prime})\omega_{n}^{-1}
\end{displaymath}
In quantum theory $A^{*}_{n}({\bf k})$ and $A_{m}({\bf
k}^{\prime})$ become creation and annihilation operators. Then,
the expectation value is
\begin{displaymath}\begin{array}{l}\frac{1}{2}\langle A^{*}_{n}({\bf
k})A_{m}({\bf k}^{\prime})+A_{m}({\bf k}^{\prime})A^{*}_{n}({\bf
k})\rangle\cr=\delta_{mn}\delta({\bf k}-{\bf
k}^{\prime})\frac{\hbar}{2}\coth(\frac{1}{2}\hbar\beta\omega_{n}).
\end{array}\end{displaymath}
When we calculate the expectation value of $\eta$ with respect to
the measure $dA\exp(-\beta {\cal H})$ then we obtain
\begin{equation}\begin{array}{l}
\langle\eta(x)\eta(x^{\prime})\rangle =\beta^{-1}\int d{\bf k}
\exp(i{\bf k}({\bf x}-{\bf
x}^{\prime}))\cr\sum_{n}\lambda_{n}^{2}\omega_{n}^{-2}
\cos(\omega_{n}(t-t^{\prime})) .\end{array}\end{equation} In
quantum theory we should consider an anticommutator on the lhs of
eq.(16). Then, $\beta^{-1}\omega_{n}^{-2}\rightarrow
\omega_{n}^{-1}\frac{1}{2}\hbar\coth(\hbar\frac{1}{2}\beta\omega_{n})$
on the rhs. So that the quantum thermal expectation values tend to
classical expectation values (in the Gibbs state) when
$\hbar\beta\rightarrow 0$.

 Let us
distinguish two regimes: small $k$ and large $k$. In the first
case we neglect ${\bf k}$ in $\omega_{n}$. Then
\begin{equation}\begin{array}{l}
\langle\eta(x)\eta(x^{\prime})\rangle
=(2\pi)^{3}\beta^{-1}\delta({\bf x}-{\bf
x}^{\prime})\cr\sum_{n}m_{n}^{-2}\lambda_{n}^{2}
\cos(m_{n}(t-t^{\prime})) .\end{array}\end{equation} Let
$(2\pi)^{\frac{3}{2}}\lambda_{n}=\tilde{\gamma} m_{n}$ then
\begin{equation}
\langle\eta(x)\eta(x^{\prime})\rangle =\tilde{\gamma}^{2}
\beta^{-1}\delta({\bf x}-{\bf x}^{\prime})\sum_{n}
\cos(m_{n}(t-t^{\prime})).
\end{equation}
If $m_{n}$ are uniformly distributed then (because of oscillations
of the cosine) the sum (18) is concentrated at small
$t-t^{\prime}$. We could e.g. apply the Poisson summation formula
(for $m_{n}\simeq n$)
\begin{displaymath}
\frac{2}{L}\sum\cos(2\pi\frac{ns}{L})=\sum_{n}\delta(s-nL)
\end{displaymath}
or approximate the sum in eq.(18) by an integral over $m_{n}$
leading to the $\delta$ function. In both cases (for a large $L$
or in the approximation
$\sum_{n}\tilde{\gamma}^{2}...=\gamma^{2}\int...$) we obtain
\begin{equation}
\langle\eta(x)\eta(x^{\prime})\rangle
=\beta^{-1}\gamma^{2}\delta({\bf x}-{\bf
x}^{\prime})\delta(t-t^{\prime})
\end{equation}with a certain $\gamma\simeq \tilde{\gamma}$.

 Next let us consider the friction $\delta \phi$ in eqs.(7) and (13). Writing
\begin{displaymath}
\sin(\omega_{n}(t-s))\omega_{n}^{-1}=\partial_{s}(\cos(\omega_{n}(t-s))\omega_{n}^{-2})
\end{displaymath}
and integrating by parts we get the term \begin{displaymath}
U^{\prime}\sum\lambda_{n}^{2}\omega_{n}^{-2}\int d{\bf k}U(t,{\bf
k})\exp(i{\bf kx})
\end{displaymath}
which under the assumptions leading to eq.(19)can be considered as
modification of the potential $V\rightarrow V+qU^{2}$ with
$q\simeq \delta(0)$ (this is the mass renormalization if $U\simeq
\phi$ as in the Leggett model \cite{leggett}). The second term
from the integration by parts
 \begin{displaymath}
-U^{\prime}\sum\lambda_{n}^{2}\omega_{n}^{-2}\int d{\bf
k}\exp(i{\bf kx})\cos(\omega_{n}t)U(0,{\bf k})
\end{displaymath}
under the assumptions used at eq.(19) is proportional to
$\delta(t)$ and to  the initial values of the field. We will
neglect this term for  $t> 0$. After the integration by parts the
friction term takes the form
\begin{displaymath}
U^{\prime}\sum_{n}\lambda_{n}^{2}\int
\omega_{n}^{-2}\cos(\omega_{n}(t-s))\partial_{s}Uds.
\end{displaymath}
We have got the same kernel as the one in eq.(16) which we
approximated by $\delta(t-s)$ in eq.(19) for the correlation of
the noise. In fact these kernels are  related by a version of  the
fluctuation dissipation theorem \cite{fd}. With these
approximations eq.(2) in Minkowski space reads
\begin{equation}
\partial_{t}^{2}\phi-\triangle\phi+V^{\prime}+\gamma^{2}U^{\prime}\partial_{t}U=U^{\prime}\eta
\end{equation}
with the noise (19). This is the equation which  would have been
derived in \cite{berera} and \cite{adv} when $a=1$. Note that in
the limit of a strong friction we obtain
\begin{displaymath}
U(\phi(t,{\bf x}))\simeq \gamma^{-2}\int^{t}\eta(s,{\bf x})ds.
\end{displaymath}
Hence, $U(\phi)$ behaves like a Brownian motion in classical field
theory theory and in the high temperature limit of the quantum
field theory. If $V^{\prime}$ is of the same order as
$\gamma^{2}U^{\prime}\partial_{t}U$ (this may be the case because
of the $U^{2}$ renormalization mentioned below eq.(19)) then
eq.(20) after the neglect of the second order derivatives reads
\begin{equation}
\gamma^{2}U^{\prime}\partial_{t}U+V^{\prime}=U^{\prime}\eta.
\end{equation}
 $U$
has a large time asymptotic distribution
\begin{displaymath}
dU\exp\Big(-\beta\gamma^{-4}\int d{\bf x}\int
dU\frac{V^{\prime}}{U^{\prime}}(U)\Big),
\end{displaymath}
where $\int dU\frac{V^{\prime}}{U^{\prime}}(U)$ means that we
first express $V^{\prime}U^{\prime -1}$ as a function of $U$ and
subsequently calculate its indefinite integral. So,
$U^{\prime}\neq 0$ may change the form of the asymptotic behaviour
of the inflaton modifying the discussion in \cite{habaimpd}.

 We
can derive the slow roll conditions leading to the approximation
(21). They require $\vert
\partial_{t}^{2}\phi\vert<< \gamma^{2}\vert U^{\prime}\partial_{t}U\vert$ and
$\vert\partial_{t}^{2}\phi\vert<< \vert V^{\prime}\vert$. These
conditions are satisfied if\begin{displaymath} \vert
V^{\prime\prime}(U^{\prime})^{-4}-2V^{\prime}(U^{\prime})^{-5}U^{\prime\prime}\vert<<
\gamma^{4}.
\end{displaymath}
The slow roll approximation can be applied to quantum fields as
well. We can see that with the friction term the quantum field
behaves like a diffusion process. In a linear model when
$V=\frac{m^{2}}{2}\phi^{2}$ and $U^{\prime}=1$ then the slow roll
requirement is $m^{2}<<\gamma^{4}$ which does not depend on
$\phi$. The  correlation functions in this model
\begin{displaymath}
\langle\phi_{t}\phi_{t^{\prime}}\rangle\simeq
\exp(-\frac{m^{2}}{\gamma^{2}}\vert t-t^{\prime}\vert)
\end{displaymath} are the same as the ones for the Ornstein-Uhlenbeck process
\cite{chandra}. The model (21) can be considered as a  limit $H<<
\gamma^{2}$ of the model in de Sitter space discussed in the next
section (in this sense $\partial_{t}U$ brings the friction in the
Minkowski space as the Hubble constant does in the de Sitter
space, in both cases the strong friction approximation is
applicable).

 Next, assume that $k$ is large in comparison
to $m_{n}$. Then, neglecting the masses in eq.(16) we have
\begin{equation}\begin{array}{l}
\langle\eta(x)\eta(x^{\prime})\rangle
=4\pi\beta^{-1}\kappa^{2}\cr\int dk\sin(k\vert{\bf x}-{\bf
x}^{\prime}\vert) (k\vert{\bf x}-{\bf x}^{\prime}\vert)^{-1}
\cos(k(t-t^{\prime}))\end{array}
\end{equation}where
\begin{displaymath}
\kappa^{2}=\sum_{n}\lambda_{n}^{2}.
\end{displaymath}

If $\vert{\bf x}-{\bf x}^{\prime}\vert=0$ then
\begin{equation}
\langle\eta(x)\eta(x^{\prime})\rangle
=8\kappa^{2}\pi^{2}\beta^{-1}\delta(t-t^{\prime})
\end{equation}
In general, from the integral (\cite{grad}, formula 3.721)
\begin{equation}
\int_{0}^{\infty}du\frac{\sin(zu)\cos(wu)}{u}=\frac{\pi}{4}(sign(z-w)+sign(z+w))\end{equation}
 we have
\begin{equation}\begin{array}{l}
\langle \eta(t,{\bf x}) \eta(t^{\prime},{\bf x}^{\prime})\rangle
=\pi^{2}\beta^{-1}\kappa^{2}\vert{\bf x}-{\bf
x}^{\prime}\vert^{-1}\cr\Big( sign(\vert{\bf x}-{\bf
x}^{\prime}\vert+t-t^{\prime})\cr+sign(\vert{\bf x}-{\bf
x}^{\prime}\vert-t+t^{\prime})\Big)\end{array}\end{equation} Note
that the covariance (25) is vanishing for the time-like
separations
\begin{equation}
\vert{\bf x}-{\bf x}^{\prime}\vert<\vert t-t^{\prime}\vert.
\end{equation}Let us still discuss various forms of the friction
term $\delta\phi$ in eq.(7). If we do not integrate by parts then
the friction in eq.(7) has an explicitly Lorentz covariant form
\begin{displaymath}
\kappa^{2}U^{\prime}\int_{0}^{t}D(x-y)U(y)d^{4}y
\end{displaymath}
where
\begin{displaymath}
iD(x-y)=\int d^{4}k\exp(-ik(x-y))sign(k_{0})\delta(k^{2})
\end{displaymath} is the commutator function for the massless
quantum scalar field. The friction term in eq.(13) can  also be
expressed in a Lorentz invariant way as
\begin{displaymath}\begin{array}{l}
\pi^{2}\kappa^{2}U^{\prime}\int_{0}^{t} \vert {\bf x}-{\bf
y}\vert^{-1}\cr\Big(\delta(\tau-\tau^{\prime}+\vert {\bf x}-{\bf
y}\vert)-\delta(\tau-\tau^{\prime}-\vert {\bf x}-{\bf
y}\vert)\Big)U(y)dy\cr=\pi^{2}\kappa^{2}U^{\prime}
\int_{0}^{t}sign(x_{0}-y_{0})\delta((x-y)^{2})U(y)d^{4}y\end{array}
\end{displaymath}
Now, eq.(7) takes the form
\begin{equation}\begin{array}{l}
\partial_{\mu}\partial^{\mu}\phi+V^{\prime}(\phi)
+4\pi^{2}\kappa^{2}U^{\prime}\int_{0}^{t} \vert {\bf x}-{\bf
y}\vert^{-1}\cr\Big(\delta(\tau-\tau^{\prime}+\vert {\bf x}-{\bf
y}\vert)-\delta(\tau-\tau^{\prime}-\vert {\bf x}-{\bf
y}\vert)\Big)U(y)dy\cr=U^{\prime}\eta
\end{array}\end{equation}
 The
kernel in eq.(13) with $ m_{n}=0$ after an integration by parts is
\begin{displaymath}\begin{array}{l}
K(t-s,{\bf x}-{\bf x}^{\prime})=4\pi\int dk k^{2}
\cos(k(t-s))k^{-2}\cr\sin(k\vert {\bf x}
 -{\bf x}^{\prime}\vert)(k\vert{\bf x}-{\bf
x}^{\prime}\vert)^{-1}\cr=\pi^{2} \vert{\bf x}-{\bf
x}^{\prime}\vert^{-1}\Big( sign(\vert{\bf x}-{\bf
x}^{\prime}\vert+t-s)\cr+sign(\vert{\bf x}-{\bf
x}^{\prime}\vert-t+s)\Big)
\end{array}
\end{displaymath}This kernel is the same as the covariance (25) of the noise and is
vanishing for time-like separations.

The stochastic equation after an integration by parts reads
\begin{equation}\begin{array}{l}
\partial_{\mu}\partial^{\mu}\phi+V^{\prime}(\phi)\cr
+\kappa^{2}U^{\prime} \int_{0}^{\tau}ds\int d{\bf y}K(\tau-s,{\bf
x}-{\bf y})\partial_{s}U(s,{\bf y})\cr=U^{\prime}\eta
\end{array}\end{equation}
with the friction kernel $K$  and the Gaussian noise (25). That
the variance of the noise and the kernel of the friction are
related follows from a version of the fluctuation dissipation
theorem \cite{fd}. Let us note that the result (27)-(28) is exact
if all $m_{n}=0$ and if the initial values of the $\chi_{n}$
fields are distributed according to the Gibbs weight factor
$\exp(-\beta {\cal H}_{n})$(the number of fields is irrelevant).
If
\begin{equation}
\vert{\bf x}-{\bf x}^{\prime}\vert<<\vert t-t^{\prime}\vert
\end{equation}
then we can approximate the kernel $K$ by a $\delta$-function
leading to the stochastic equation
\begin{equation}
\partial_{\mu}\partial^{\mu}\phi+V^{\prime}(\phi)+2\pi^{2}\kappa^{2}U^{\prime}
\int d{\bf y}\partial_{t}U(t,{\bf y})=U^{\prime}\eta
\end{equation}
where $\eta$ in this approximation has the correlation (23). The
condition (29) can  be interpreted as a negligence of the spatial
dependence of $\phi$. If the spatial dependence of $\phi$ is
ignored then we could identify eqs.(20) and (30) although we have
derived them on a basis of different assumptions concerning the
$\chi$ fields (this could explain the "alternative" derivation of
the stochastic equation in Appendix B of \cite{berera}). In the
strong friction limit we obtain
\begin{displaymath}
2\pi^{2}\int d{\bf y}U(t,{\bf y})\simeq\kappa^{-2}\int^{t}
\eta(s)ds
\end{displaymath}
In eq.(30) we can take  $V^{\prime}$ into account and integrate
both sides over  ${\bf x}$. Then, we could conclude that $\int
d{\bf y}U$ has an equilibrium distribution. In contradistinction
to the strong friction limit of eq.(20) only the space average
tends to the Brownian motion in eq.(30) (and eventually to an
equilibrium). However, there is a distinction between the results
(20) and (30). Eq.(20) is derived under an assumption of an
infinite number of $\chi$ fields with properly chosen masses,
whereas eq.(30) follows from eq.(27) which is exact if $m_{n}=0$.
In conclusion, in the strong friction limit the classical thermal
field as well as the thermal quantum fields behave as a stochastic
process. As an example, if in eq.(30) we neglect second order time
derivatives and assume $V=U^{2}$ then
\begin{displaymath}\langle\int d{\bf y}U(t,{\bf y})\int d{\bf
y}^{\prime}U(t^{\prime},{\bf y}^{\prime}) \rangle\simeq
\exp(-\pi^{-2}\kappa^{-2}\vert t-t^{\prime}\vert)
\end{displaymath}
Hence, $\int d{\bf y}U(t,{\bf y})$ is the Ornstein-Uhlenbeck
stochastic process.

 \section{Exponentially expanding
environment} In an expanding universe we write
\begin{equation}
\chi=a^{-\frac{3}{2}}\tilde{\chi}.
\end{equation}
Then, in the momentum space\begin{equation}
\partial_{t}^{2}\tilde{\chi}_{n}+\omega_{n}^{2}\tilde{\chi}_{n}=-\lambda_{n}a^{\frac{3}{2}}\phi,
\end{equation} where
\begin{displaymath}
\omega_{n}^{2}=a^{-2}k^{2}+m_{n}^{2}-\frac{3}{2}\partial_{t}H-\frac{9}{4}H^{2}.
\end{displaymath} First, let us consider low momenta $a^{-1}k<<H$ so that for a large time
(as in \cite{adv}) we can neglect $a^{-2}k^{2} $ term. Then, we
assume that $\omega_{n}^{2}>0$, $H$ is slowly varying and
$\omega_{n} $ are approximately constant. The approximate
solutions of eq.(32) at $\lambda=0$ are (the same as in eq.(11))
\begin{equation}
\tilde{u}_{n}(k,x)=\frac{1}{\sqrt{2\omega_{n}}}(2\pi)^{-\frac{3}{2}}\exp(i{\bf
kx}-i\omega_{n} t).
\end{equation}
They are normalized as
\begin{displaymath}
\int d{\bf
x}(\tilde{u}_{n}^{\prime}\partial_{0}\tilde{u}_{n}^{*}-\tilde{u}_{n}^{*}\partial_{0}\tilde{u}_{n}^{\prime})=i\delta({\bf
k}-{\bf k}^{\prime}).
\end{displaymath}
In de Sitter space $H=const$. If we neglect $a^{-2}k^{2}$ and
assume $m_{n}^{2}>\frac{9}{4}H^{2}$ then we can apply the same
approximation as in sec.3. Now, eq.(7) in de Sitter space reads
(for $a(t)=\exp(Ht)$)
\begin{equation}\begin{array}{l}
\partial_{t}^{2}\phi-a^{-2}\triangle\phi+3H\partial_{t}\phi+V^{\prime}
+\gamma^{2}U^{\prime}\partial_{t}U\cr+\frac{3}{2}\gamma^{2}HU^{\prime}U
=a^{-\frac{3}{2}}U^{\prime}\eta
\end{array}\end{equation}
where the last two terms on the lhs of eq.(34) result from the
integration by parts as in eq.(20) with an extra term coming from
the transformation (31).The noise has the
covariance\begin{equation}
\langle\eta(x)\eta(y)\rangle=\beta^{-1}\gamma^{2}\delta(x-y)
\end{equation}
At slow roll approximation with the neglect of second order
derivatives we
 obtain
\begin{equation}\begin{array}{l}
3H\partial_{t}\phi +V^{\prime}+\gamma^{2}U^{\prime
2}\partial_{t}\phi+\frac{3}{2}\gamma^{2}HUU^{\prime}
=a^{-\frac{3}{2}}U^{\prime}\eta
\end{array}\end{equation} with $a=\exp(Ht)$.
The slow roll requirements are $\vert\partial_{t}^{2}\phi\vert<<
\vert(3H+\gamma^{2}(U^{\prime})^{2})\partial_{t}\phi\vert$ and
$\vert\partial_{t}^{2}\phi\vert<<\vert V^{\prime}
+\frac{3}{2}\gamma^{2}HU^{\prime}U\vert$. These conditions are
satisfied if
\begin{displaymath}
\vert\Big((V^{\prime}
+\frac{3}{2}\gamma^{2}HU^{\prime}U)(3H+\gamma^{2}(U^{\prime})^{2})^{-1}\Big)^{\prime}\vert
<<3H+\gamma^{2}(U^{\prime})^{2} \end{displaymath} As an example,
if $H=const$, $U^{\prime}=1$ and $V^{\prime}=m^{2}\phi$ then the
slow roll condition is  $ m^{2}<< (3H +\gamma^{2})^{2}$. It does
not depend on $\phi$ and allows to solve the slow roll stochastic
equation explicitly.If $H$ is determined by Einstein equations
then $H$ depends on $\phi$ (as will be discussed in sec. 6). Then,
the slow roll conditions still must involve some requirements
allowing to determine $H$ as a function of $\phi$.

 The solution of eq.(36)in
the limit of large $\gamma$ is
\begin{displaymath}\begin{array}{l}
\delta U\equiv U(\phi(t,{\bf
x}))-\exp(-\frac{3}{2}Ht)U(\phi(0,{\bf x}))\cr =
\gamma^{-2}\exp(-\frac{3}{2}Ht)\int_{0}^{t}\eta(s,{\bf x})ds
\end{array}\end{displaymath}
The strong friction covariance  $\langle \delta U_{t}({\bf
x})\delta U_{t^{\prime}}({\bf x}^{\prime})\rangle$ is $\exp(
-3Ht-3Ht^{\prime})min(t,t^{\prime})\beta^{-1}\gamma^{-2}\delta({\bf
x}-{\bf x}^{\prime})$ . Hence, we get an extra damping factor in
comparison to the case  of the Minkowski space (discussed at the
end of sec.3).

It is instructive to consider the soluble case of eq.(36) with
$U^{\prime}=1$, $V^{\prime}=m^{2}\phi$. Then, the solution (with
zero initial condition) is the Ornstein-Uhlenbeck process
\begin{equation} \phi_{t}
=\exp(-\frac{3}{2}Ht)\int_{0}^{t}\exp(-\frac{m^{2}}{3H+\gamma^{2}}s)\eta_{s}ds
\end{equation}
We have got the same result as the one for the de Sitter space
with the Hubble constant $H\rightarrow H+\frac{\gamma^{2}}{3}$ or
in Minkowski space with a friction (and $H=0$).

  Another (complementary)
regime is the one when the term $a^{-2}k^{2}$ is dominating
($a^{-1}k>>H$ and $a^{-1}k>>m_{n}$). In such a case
$\omega_{n}\simeq a^{-1}k$. Hence, in the solution (33) of the
wave equation (32) ($\lambda_{n}=0$)
\begin{displaymath} \exp(i\int_{0}^{t}\omega_{n})\simeq \exp(i
k\int_{0}^{t}a^{-1})= \exp(ik\tau),
\end{displaymath}
where $\tau$ is the conformal time.
 In the conformal time (for the exponential expansion $a(t)=\exp(Ht)$)
\begin{displaymath}
ds^{2}=(H\tau)^{-2}(d\tau^{2}-d{\bf x}^{2}).
\end{displaymath}
eq.(5) reads
\begin{equation}\begin{array}{l}
\partial_{\tau}^{2}\chi_{n}-\triangle\chi_{n}-2\tau^{-1}\partial_{\tau}\chi_{n}
+m_{n}^{2}(\tau H)^{-2}\chi_{n}\cr=-\lambda_{n}(H\tau)^{-2}U
\end{array}\end{equation} and eq.(2)

\begin{equation}\begin{array}{l}
\partial_{\tau}^{2}\phi-\triangle\phi-2\tau^{-1}\partial_{\tau}\phi\cr
+V^{\prime}(\tau
H)^{-2}=-U^{\prime}\sum_{n}\lambda_{n}(H\tau)^{-2}\chi_{n}
\end{array}\end{equation}
The free Lagrangian of the $\chi$ fields is
\begin{equation}
L=\frac{1}{2}(H^{2}\tau^{2}(\partial_{\tau}\chi)^{2}-H^{2}\tau^{2}(\nabla\chi)^{2}-m^{2}\chi^{2})
\end{equation}
The statistical weight (9) with the Hamiltonian ${\cal H}=T^{00}$
 (with the momentum $\Pi=\tau^{2}(\partial_{\tau}\chi$)) is
\begin{equation}\begin{array}{l}
\exp(-\int_{0}^{\beta}d\tau d{\bf x}\sqrt{g}{\cal
H})\cr=\exp\Big(-\frac{1}{2}\int_{0}^{\beta}d\tau\int d {\bf
x}(H\tau)^{-4}(H^{2}\tau^{2}(\partial_{\tau}\chi)^{2}\cr+H^{2}\tau^{2}(\nabla\chi)^{2}+m^{2}\chi^{2})\Big)
\end{array}\end{equation} For a large $k$ we may ignore the masses $m_{n}$.
Then, the solution of the free wave equation is\begin{equation}
\tau^{\frac{3}{2}}H^{(1)}_{\frac{3}{2}}(k\tau)\simeq
k^{-\frac{1}{2}}\tau\exp(ik\tau)(1+\frac{i}{k\tau})
\end{equation}where $H^{(1)}_{\frac{3}{2}}$ is the Hankel function
\cite{grad}.

 We can write the general solution in the form
\begin{equation}\begin{array}{l}\chi^{cl}= \tilde{A}({\bf
k})k^{-\frac{1}{2}}\tau\exp(ik\tau)(1+\frac{i}{k\tau})\cr+\tilde{A}^{*}({\bf
k})k^{-\frac{1}{2}}\tau\exp(-ik\tau)(1-\frac{i}{k\tau})\end{array}
\end{equation}If we insert the solution (43) in the Hamiltonian (41) and neglect the terms (decaying fast
for large $\tau$;large $\tau$ means small $a$, hence close to the
Big Bang)
\begin{equation}
\tau^{-2}\int d{\bf k}k^{-1}\tilde{A}({\bf k})^{2}\exp(2ik\tau)
\end{equation}
and $\tau^{-2}\int (\partial_{\tau}\chi)^{2}$ ,then
\begin{equation}\int d{\bf x}
\sqrt{g}{\cal H}\simeq H^{-2}\int d{\bf k}\vert {\bf
k}\vert\tilde{A}^{*}({\bf k})\tilde{A}({\bf k})
\end{equation} The $\tau$-dependence  cancels for a large $\tau$
in eq.(45) as expressed  in eq.(41) because $\tau^{-2}$ in
$\sqrt{g}{\cal H}$ cancels with $\tau^{2}$ in $(\nabla \chi)^{2}$
resulting from the terms $\tau\exp(ik\tau)$  in eq.(43).

We can calculate the expectation value  of the noise
\begin{equation}\begin{array}{l} \eta(\tau,{\bf x })=(H\tau)^{-2}\int d{\bf k}\exp(i{\bf
kx})\cr\sum_{n}\lambda_{n}k^{-\frac{1}{2}}(\tilde{A}_{n}({\bf
k})\tau\exp(ik\tau)(1+\frac{i}{k\tau})\cr+\tilde{A}_{n}^{*}({\bf
k})\tau\exp(-ik\tau)(1-\frac{i}{k\tau})))
\end{array}\end{equation} with respect
to the weight factor (41). We obtain
\begin{equation}\begin{array}{l}\langle
\eta(x)\eta(x^{\prime})\rangle=(H\tau)^{-1}(H\tau^{\prime})^{-1}\beta^{-1}\kappa^{2}
4\pi\int dk k\cr\Big(\frac{\cos(k(\tau-\tau^{\prime}))}{k}
\cr-\frac{\sin(k(\tau-\tau^{\prime}))}{k^{2}\tau}+\frac{\sin(k(\tau-\tau^{\prime}))}{k^{2}\tau^{\prime}}
+\frac{\cos(k(\tau-\tau^{\prime}))}{k^{3}\tau\tau^{\prime}}\Big)\cr
\sin(k\vert {\bf x}-{\bf y}\vert)(k\vert {\bf x}-{\bf
y}\vert)^{-1}
\end{array}\end{equation} We can see that the last term of the sum in eq.(47) is divergent at small
$k$. This is a typical infrared divergence of a massless $\chi$
field theory  in de Sitter space. It means that instead of the
random variable $\eta$ we should consider $\eta(x)-\eta(z)$ with a
fixed $z=(s,{\bf z})$. Then, the variance

\begin{displaymath}\begin{array}{l}\langle
(\eta(x)-\eta(z))^{2})\rangle=H^{-2}\beta^{-1}\kappa^{2} 4\pi\int
dk k\Big(\frac{1}{k}(\tau^{-2}+s^{-2}) \cr-2(\tau
s)^{-1}\cos(k(\tau-s))\sin(k\vert {\bf x}-{\bf z}\vert)(k^{2}\vert
{\bf x}-{\bf z}\vert)^{-1}
\cr+(2\frac{\sin(k(\tau-s))}{k^{2}\tau^{2}s}-2\frac{\sin(k(\tau-s))}{k^{2}s^{2}\tau})\sin(k\vert
{\bf x}-{\bf z}\vert)(k\vert {\bf x}-{\bf z}\vert)^{-1}
\cr+\frac{\tau^{4}+s^{4}}{k^{3}\tau^{4}s^{4}}\cr
-2\tau^{2}s^{2}\cos(k(\tau-s))\frac{1}{k^{3}\tau^{4}s^{4}}\sin(k\vert
{\bf x}-{\bf z}\vert)(k\vert {\bf x}-{\bf z}\vert)^{-1}\Big)
\end{array}\end{displaymath}
is finite.

 Using the formula
(24) we calculate the Fourier transform  in eq.(47) (neglecting
the terms decaying fast for a large time) with the result
\begin{equation}\begin{array}{l}
\langle \eta(\tau,{\bf x}) \eta(\tau^{\prime},{\bf y})\rangle
=(H\tau)^{-1}(H\tau^{\prime})^{-1}\beta^{-1}\kappa^{2}\pi^{2}\cr
\vert{\bf x}-{\bf y}\vert^{-1}\Big( sign(\vert{\bf x}-{\bf
y}\vert+\tau-\tau^{\prime})\cr+sign(\vert{\bf x}-{\bf
y}\vert-\tau+\tau^{\prime})\Big)\end{array}\end{equation} When $
\vert{\bf x}-{\bf y}\vert\rightarrow 0$ then in eq.(48) we obtain
$\delta(\tau-\tau^{\prime})$ (this can be seen either from eq.(47)
or from eq.(48) by a formal Taylor expansion in $ \vert{\bf
x}-{\bf y}\vert$ ).

 The solution of eq.(38) with $m_{n}=0$ (at large $k$ we neglect the masses) can be expressed by the Green function
 ${\cal G}$ (discussed in detail in the next section, eq.(73))
 \begin{equation}\begin{array}{l}
\chi_{n}=-\lambda_{n}\tau\int_{\tau}^{\infty}{\cal
G}(\tau,\tau^{\prime})\tau^{\prime-1}(\tau^{\prime}H)^{-2}U
d\tau^{\prime} +\chi_{n}^{cl} \end{array}
\end{equation}where $\chi^{cl}$ is the solution with
$\lambda_{n}=0$ and
\begin{displaymath}
\begin{array}{l} {\cal G}(\tau,\tau^{\prime})
=-\Big(\frac{\sin(k(\tau-\tau^{\prime}))}{k}
\cr+\frac{\cos(k(\tau-\tau^{\prime}))}{2k^{2}\tau}-\frac{\cos(k(\tau-\tau^{\prime}))}{2k^{2}\tau^{\prime}}
+\frac{\sin(k(\tau-\tau^{\prime}))}{k^{3}\tau\tau^{\prime}}\Big)
\end{array}\end{displaymath}
 Using the formulas 3.741 of \cite{grad}
\begin{equation}
\int_{0}^{\infty}du\frac{\sin(zu)}{u}=\frac{\pi}{2}sign(z)
\end{equation}

\begin{displaymath}
\int_{0}^{\infty}du\frac{\sin(zu)\sin(wu)}{u^{2}}=\frac{\pi}{2}min(w,z)
\end{displaymath}and the integral (24) we can calculate the Fourier integrals
in order to express the kernel ${\cal G}$ in configuration space
\begin{equation}\begin{array}{l} {\cal G}(\tau,{\bf x};\tau^{\prime},{\bf y})
=4\pi^{2} \vert {\bf x}-{\bf
y}\vert^{-1}\cr\Big(\delta(\tau-\tau^{\prime}+\vert {\bf x}-{\bf
y}\vert)-\delta(\tau-\tau^{\prime}-\vert {\bf x}-{\bf y}\vert)\cr
+\frac{1}{8}(\tau^{-1}-\tau^{\prime-1})(sign(\vert {\bf x}-{\bf
y}\vert+\tau-\tau^{\prime})\cr+sign(\vert {\bf x}-{\bf
y}\vert-\tau+\tau^{\prime}))
+\frac{1}{4\tau\tau^{\prime}}min(\tau-\tau^{\prime},\vert {\bf
x}-{\bf y}\vert)\Big)
\end{array}\end{equation}
Note that\begin{displaymath}\begin{array}{l} \vert {\bf x}-{\bf
x}^{\prime}\vert^{-1}\Big(\delta (\tau-\tau^{\prime}-\vert {\bf
x}-{\bf x}^{\prime}\vert)-\delta (\tau-\tau^{\prime}+\vert {\bf
x}-{\bf x}^{\prime}\vert)\Big)\cr =2
sign(x_{0}-x_{0}^{\prime})\delta(
(x-x^{\prime})^{2})\end{array}\end{displaymath} where
$(x-x^{\prime})^{2}$ is the Minkowski distance. Hence, the leading
term (for a large time) is Lorentz invariant.

 After an insertion of the solution of eq.(39) in eq.(38) the stochastic equation (7) for $\phi$ in
the configuration space reads
\begin{equation}\begin{array}{l}
(\partial_{\tau}^{2}\phi-\triangle\phi-2\tau^{-1}\partial_{\tau})\phi
+(\tau H)^{-2}V^{\prime}\cr=\kappa^{2}(H\tau)^{-2} \tau
U^{\prime}\int_{\tau}^{\infty}{\cal
G}(\tau,\tau^{\prime})\tau^{\prime-1}(\tau^{\prime}H)^{-2}Ud\tau^{\prime}
\cr+U^{\prime}(\tau H)^{-2}\eta \end{array}
\end{equation}

 For  $\vert {\bf
x}-{\bf y}\vert << \vert \tau-\tau^{\prime}\vert$ and large $\tau$
and $\tau^{\prime}$ we get the stochastic
equation\begin{equation}\begin{array}{l}
\partial_{\tau}^{2}\phi-\triangle\phi-2\tau^{-1}\partial_{\tau}\phi
+(\tau H)^{-2}V^{\prime}\cr+8\pi^{2}\kappa^{2}
U^{\prime}(H\tau)^{-4}\int d{\bf
y}\partial_{\tau}U\cr+24\pi^{2}\kappa^{2}\tau^{-5}H^{-4}U^{\prime}\int
d{\bf y}U=U^{\prime}(H\tau)^{-2}\eta\end{array}
\end{equation} by an expansion
of the kernel (51) in $\vert {\bf x}-{\bf y}\vert $ and
integration by parts over $\tau^{\prime}$ in eq.(52) (as in an
analogous derivation of eq.(30)). The limit of small $\vert {\bf
x}-{\bf y}\vert $ of the noise follows from eq.(48)

\begin{equation}\begin{array}{l}
\langle \eta(\tau,{\bf x}) \eta(\tau^{\prime},{\bf y})\rangle =2(H
\tau)^{-2}\pi^{2}\kappa^{2}\beta^{-1}
  \delta(
\tau-\tau^{\prime})\cr=2\pi^{2}\kappa^{2}\exp(-3Ht)\beta^{-1}
  \delta(
t-t^{\prime})\end{array}\end{equation} In the cosmic time eq.(53)
reads
\begin{displaymath}\begin{array}{l}
\partial_{t}^{2}\phi-\exp(-2Ht)\triangle\phi+3H\partial_{t}\phi
+V^{\prime}\cr+8\pi^{2}\kappa^{2} \exp(3Ht)U^{\prime}\int d{\bf
y}\partial_{t}U\cr+24\kappa^{2}\pi^{2}H\exp(3Ht)U^{\prime}\int
d{\bf y}U=U^{\prime}\eta\end{array}
\end{displaymath} With the  neglect of $\partial_{t}^{2}\phi$
(slow-roll) and $ \triangle \phi$ ( which disappears after a space
average in eq.(53)) we obtain

\begin{equation}\begin{array}{l}
\exp(-3Ht)(3H\partial_{t}\phi +V^{\prime})+8\pi^{2}\kappa^{2}
U^{\prime}\int d{\bf y}\partial_{t}U\cr+24\kappa^{2}\pi^{2}H
U^{\prime}\int d{\bf y}U=U^{\prime}\exp(-3Ht)\eta\end{array}
\end{equation}
In the limit of the strong friction (only $\kappa^{2}$ terms on
the lhs)

\begin{displaymath}\begin{array}{l}\partial_{t}\int d{\bf y}U+3H\int
d{\bf y}U=\frac{1}{8\pi^{2}\kappa^{2}}\exp(-3Ht)\eta\end{array}
\end{displaymath}
The solution is an Ornstein-Uhlenbeck process
\begin{displaymath}\begin{array}{l}
\int d{\bf y}U_{t}=\exp(-3Ht)\int d{\bf
y}U_{0}\cr+\frac{1}{8\pi^{2}\kappa^{2}}\exp(-3Ht)\int_{0}^{t}\eta(s)ds
\end{array}\end{displaymath} such that
\begin{displaymath}\begin{array}{l}
\langle (\int d{\bf y}U_{t}-\langle\int d{\bf y}U_{t}\rangle)
(\int d{\bf y}^{\prime}U_{t^{\prime}}-\langle\int d{\bf
y}^{\prime}U_{t^{\prime}}\rangle)
\rangle\cr=\frac{1}{32\pi^{2}\kappa^{2}}
\frac{1}{3H\beta}\exp(-3Ht-3Ht^{\prime})\Big(1-\exp(-3H
min(t,t^{\prime}))\Big)
\end{array}\end{displaymath}
We can see that the classical field in a thermal environment tends
to a diffusion process. We could prove the same behaviour for the
quantum field in a thermal state. It seems that in both cases the
appearance of fluctuations and friction is crucial for the
diffusive behaviour. This phenomenon is similar to the diffusive
behaviour of zero temperature quantum fields in de de Sitter space
which also results from "friction" $H\partial_{t}\phi$ and from
quantum fluctuations \cite{star}\cite{vilenkin3}\cite{staryoko}.

\section{Environment
in an almost exponential expansion}
 We consider a flat expanding metric which is close to
 de Sitter.
For $k<<aH$ we may repeat the calculations of sec.4 and
ref.\cite{adv} leading to an analog of the stochastic equation
(34) (in the cosmic time)
\begin{equation}\begin{array}{l}
g^{-\frac{1}{2}}\partial_{\mu}g^{\frac{1}{2}}g^{\mu\nu}\partial_{\nu}\phi
+V^{\prime}+\gamma^{2}U^{\prime}\partial_{t}U+
\frac{3}{2}\gamma^{2}HU^{\prime}U\cr=U^{\prime}g^{-\frac{1}{4}}\eta
\end{array}\end{equation}with the noise (19). It can be shown that
without the friction terms eq.(56) is invariant with respect to
the general change of coordinates. With the friction terms even
the Lorentz invariance is violated because in the derivation of
eq.(56) from eqs.(2)-(3) we have neglected the spatial derivatives
of $\chi$ fields.

 Next, we explore $\chi$ fields with large
$k>>aH$
 in a conformal  time (we follow to some extent
 our earlier paper \cite{habaepj2})
 \begin{displaymath}
 ds^{2}=a^{2}(d\tau^{2}-d{\bf x}^{2}),
 \end{displaymath}
where the conformal time $\tau$ is
\begin{displaymath}
\tau=\int dt a^{-1}.
\end{displaymath}With a slowly varying $H$ we have approximately
\begin{equation}
aH=-(1-\epsilon)^{-1}\frac{1}{\tau},
\end{equation}
where
\begin{equation}
\epsilon=-H^{-2}\partial_{t}H
\end{equation}
($t$ is the cosmic time).

If the expansion is close to exponential then  with the relation
(57) eq.(3) is
\begin{equation}\begin{array}{l}
(\partial_{\tau}^{2}-\frac{2}{1-\epsilon}\frac{1}{\tau}\partial_{\tau}+k^{2}
+\frac{3\eta_{n}}{(1-\epsilon)^{2}}\tau^{-2})\chi_{n}\cr
=-\lambda_{n} (H\tau)^{-2}(1-\epsilon)^{-2}U,\end{array}
 \end{equation}
where
\begin{displaymath}
3\eta_{n}=m_{n}^{2}H^{-2}
\end{displaymath}
Let
\begin{equation}
\chi_{n}=\tau^{\alpha}\Psi_{n}
\end{equation}
with
\begin{equation}
\alpha=\frac{1}{1-\epsilon}.
\end{equation}
Then
\begin{equation}\begin{array}{l}
(\partial_{\tau}^{2}+k^{2}
+\frac{-2+3\eta_{n}+\epsilon}{(1-\epsilon)^{2}}\tau^{-2})\Psi_{n}\cr=-\lambda_{n}
\tau^{-\alpha} (H\tau)^{-2}(1-\epsilon)^{-2}U.\end{array}
 \end{equation}
 We obtain another form of eq.(3) if
 we write \begin{equation}
 \chi_{n}=\tau^{\mu}\tilde{\Psi}_{n}
 \end{equation}
 with
\begin{equation}
\mu=(1-\epsilon)^{-1}(\frac{3}{2}-\frac{\epsilon}{2}).
\end{equation}
Then
\begin{equation}\begin{array}{l}
(\partial_{\tau}^{2}+\tau^{-1}\partial_{\tau}+k^{2}-\nu_{n}^{2}\tau^{-2})\tilde{\Psi}_{n}\cr=
-\tau^{-\mu}\lambda_{n} (H\tau)^{-2}(1-\epsilon)^{-2}U
\end{array}\end{equation} where
\begin{equation}
\nu_{n}^{2}=(1-\epsilon)^{-2}\Big((\frac{3}{2}-\frac{\epsilon}{2})^{2}-3\eta_{n}\Big).
\end{equation}
With $\lambda_{n}=0$ the solution of eq.(65) is the Hankel
function \cite{grad} $\tilde{\Psi}_{n}=H_{\nu_{n}}^{(1)}(k\tau)$.
For a general expanding metric the solution of a homogeneous
eq.(59) ($\lambda_{n}=0$) is
\begin{displaymath}
 u_{n}=\tau^{\mu}H^{(1)}_{\nu_{n}}(k\tau).
\end{displaymath}  In
general, we have a superposition of classical solutions with
different $k$
\begin{equation}
\chi^{cl}_{n}=
A_{n}(k)u_{n}(k)+A^{*}_{n}(k)u^{*}_{n}(k).\end{equation}
 The noise is
\begin{equation}
\eta=-(H\tau(1-\epsilon))^{-2}\sum_{n}\chi_{n}^{cl}
\end{equation}

The solution of eq.(62) with $\lambda_{n}=0$ on the rhs
 is $\psi_{\nu}=\tau^{-\alpha+\mu}H_{\nu}$. Then, the solution
of the  wave equation (62) for $\Psi_{n}$ with $U(\phi)$ on the
rhs is
\begin{equation}\begin{array}{l}
\Psi_{n}(\tau)=-\lambda_{n}\int_{\tau}^{\infty} {\cal
G}(\tau,\tau^{\prime})\tau^{\prime-\alpha}(H\tau^{\prime}(1-\epsilon))^{-2}U(\tau^{\prime})d\tau^{\prime}
\end{array}\end{equation} where ${\cal G}$ is the Green function
of the operator on the lhs of eq.(62). The Green function can be
constructed from the two independent solutions of the homogeneous
equation (62)

\begin{equation}
\psi_{1}(k\tau)=(k\tau)^{-\alpha+\mu}J_{\nu}(k\tau),
\end{equation}\begin{equation}
\psi_{2}(k\tau)=(k\tau)^{-\alpha+\mu}Y_{\nu}(k\tau),
\end{equation} where from eqs.(61) and (64)
\begin{equation}
\mu-\alpha=\frac{1}{2}.
\end{equation}
 The Bessel functions $J$ and $Y$ \cite{grad} are defined by the Hankel function
$H_{\nu}^{(1)}=J_{\nu}+iY_{\nu}$. The Green function for
$\tau<\tau^{\prime}$ is \begin{equation} {\cal G}
(\tau,\tau^{\prime})=w(k)^{-1}(\psi_{1}(k\tau)\psi_{2}(k\tau^{\prime})-\psi_{2}(k\tau)\psi_{1}(k\tau^{\prime})),
\end{equation}where $w(k)$ is the wronskian. If $a$ is close to exponential then  $w(k)=\frac{2k}{\pi}$.  The Green function could also be expressed as
\begin{equation}\begin{array}{l}(k^{2}\tau \tau^{\prime})^{\frac{1}{2}}\frac{\pi}{4ik}(H_{\nu}^{(1)*}(k\tau)
H_{\nu}^{(1)}(k\tau^{\prime})-H_{\nu}^{(1)*}(k\tau^{\prime})
H_{\nu}^{(1)}(k\tau))\end{array}
\end{equation}We insert the Green function (74) in eq.(69) in order to
calculate $\delta\phi$ of eq.(7) for a large time (we use the
asymptotic expansion of $H_{\nu}$ \cite{grad}).Then
\begin{equation}\begin{array}{l}
\delta\phi_{\tau}({\bf
x})=\kappa^{2}(H\tau(1-\epsilon))^{-2}\tau^{\alpha}U^{\prime}\cr\int_{\tau}^{\infty}
{\cal
G}(\tau,\tau^{\prime})\tau^{\prime-\alpha}(H\tau^{\prime}(1-\epsilon))^{-2}U(\tau^{\prime},{\bf
k})d\tau^{\prime} \exp(i{\bf kx})d{\bf
k}\cr\simeq\kappa^{2}(H\tau(1-\epsilon))^{-2}
\tau^{\alpha}U^{\prime}\int_{\tau}^{\infty}
k^{-1}\sin(k(\tau-\tau^{\prime}))
\tau^{\prime-\alpha}\cr(H\tau^{\prime}(1-\epsilon))^{-2}U(\tau^{\prime},{\bf
k})d\tau^{\prime}\exp(i{\bf kx})d{\bf k}\cr\equiv
\kappa^{2}U^{\prime}\int \tilde{K}({\tau,\tau^{\prime};\bf
k})U(\tau^{\prime},{\bf k})d\tau^{\prime}\exp(i{\bf kx})d{\bf k}
\end{array}\end{equation}
where
\begin{equation}\begin{array}{l}\tilde{K}(\tau,\tau^{\prime};{\bf
x}-{\bf y})=\kappa^{2}(H\tau(1-\epsilon))^{-2}\tau^{\alpha}{\cal
G}(\tau,\tau^{\prime};{\bf x}-{\bf y})\cr
\tau^{\prime-\alpha}(H\tau^{\prime}(1-\epsilon))^{-2}.
\end{array}\end{equation}
From eq.(75)  there follows the stochastic equation (7)
\begin{equation}\begin{array}{l}
(\partial_{\tau}^{2}\phi-\triangle\phi-2\tau^{-1}\partial_{\tau})\phi
+V^{\prime}(\tau(1-\epsilon)H)^{-2}\cr=U^{\prime}\kappa^{2}(H\tau(1-\epsilon))^{-2}
\tau^{\alpha}\cr\int_{\tau}^{\infty}{\cal
G}(\tau,\tau^{\prime})\tau^{\prime-\alpha}(\tau^{\prime}H(1-\epsilon))^{-2}Ud\tau^{\prime}
+U^{\prime}\eta .\end{array}
\end{equation}

 In eq.(77)  we have a
non-Markovian and non-local friction term. For a large time
 ${\cal G}$ can be approximated by the first term of eq.(51). If we assume that only small $\vert {\bf x}-{\bf y}\vert$
contribute and expand in this variable then
\begin{equation}\begin{array}{l}-U^{\prime}(H\tau(1-\epsilon))^{-2}\int d\tau^{\prime}
\tau^{\alpha}
\tau^{\prime-\alpha}(H\tau^{\prime}(1-\epsilon))^{-2} \vert {\bf
x}-{\bf y}\vert^{-1}\cr \Big(\delta (\tau-\tau^{\prime}-\vert {\bf
x}-{\bf y}\vert)-\delta (\tau-\tau^{\prime}+\vert {\bf x}-{\bf
y}\vert)\Big)U(\tau^{\prime},{\bf y})d{\bf y} \cr \simeq
2(2+\alpha)\tau^{-5}(H(1-\epsilon))^{-4}\int d{\bf y}U(\tau,{\bf
y})\cr+2(H\tau(1-\epsilon))^{-4}U^{\prime}\int d{\bf
y}\partial_{\tau}U(\tau,{\bf y}).
\end{array}\end{equation}
As a result for a small $\vert {\bf x}-{\bf y}\vert$ we obtain a
stochastic equation generalizing eq.(53)

\begin{equation}\begin{array}{l}
\partial_{\tau}^{2}\phi-\triangle\phi-2\tau^{-1}\partial_{\tau}\phi
+(\tau(1-\epsilon) H)^{-2}V^{\prime}\cr+8\pi^{2}\kappa^{2}
U^{\prime}(H(1-\epsilon)\tau)^{-4}\int d{\bf
y}\partial_{\tau}U\cr+8(3+\epsilon)\pi^{2}\kappa^{2}\tau^{-5}((1-\epsilon)H)^{-4}U^{\prime}\int
d{\bf y}U\cr =U^{\prime}\eta\end{array}
\end{equation}
The covariance of the noise is equal to an expectation value with
respect to the measure (9)
\begin{equation}\begin{array}{l}
\langle
\eta(x)\eta(x^{\prime})\rangle=(H(1-\epsilon)\tau)^{-2}(H(1-\epsilon)\tau^{\prime})^{-2}\kappa^{2}\cr
\int d{\bf k}d{\bf k}^{\prime}\exp(i{\bf k}{\bf x}-i{\bf
k}^{\prime}{\bf x}^{\prime})\langle A_{m}^{*}({\bf k})A_{n}({\bf
k}^{\prime})\cr \tau^{\mu-\frac{1}{2}}\tau^{\prime
\mu-\frac{1}{2}}(H_{\nu}^{(1)*}(k\tau)
H_{\nu}^{(1)}(k\tau^{\prime})+H_{\nu}^{(1)*}(k\tau^{\prime})
H_{\nu}^{(1)}(k\tau))\rangle.\end{array}
\end{equation}
where ( as $m_{n}=0$ ) \begin{equation}
\nu=\mu=\frac{\frac{3}{2}-\frac{\epsilon}{2}}{1-\epsilon}
\end{equation}
The asymptotic behaviour  of the product of Hankel functions in
eq.(80) (the leading term) does not depend on $\nu$. Hence,the
approximation (45) still holds true. As a consequence for a large
time we obtain

\begin{equation}\begin{array}{l}
\langle \eta(\tau,{\bf x}) \eta(\tau^{\prime},{\bf y})\rangle
=(H(1-\epsilon)\tau)^{-2}(H(1-\epsilon)\tau^{\prime})^{-2}\cr
H^{2}\pi^{2}\kappa^{2} \tau^{\mu-\frac{1}{2}}\tau^{\prime \mu
-\frac{1}{2}}  \vert{\bf x}-{\bf y}\vert^{-1}\cr\Big(
sign(\vert{\bf x}-{\bf y}\vert+\tau-\tau^{\prime})+sign(\vert{\bf
x}-{\bf y}\vert-\tau+\tau^{\prime})\Big)\end{array}\end{equation}
In an expanding universe we cannot expect a stationary evolution
of the inflaton if $H$ is almost  a constant. However, if in the
stochastic equations derived in this paper we insert the formula
for $H$ in terms of the potentials which result from  the
slow-roll approximation then we can consider again the time
evolution close to the slow-roll as discussed in \cite{invmeasure}
but now with the modified stochastic equation as briefly discussed
in the next section.

\section{Einstein-Klein-Gordon system}
In this section we are going to determine the Hubble variable $H$
and the expansion scale $a$  as solutions of Einstein equations.
We consider the Lagrangian (1) as a source of the energy momentum
for Einstein equations
\begin{equation} G_{\mu\nu}=8\pi GT_{\mu\nu}
\end{equation}
where $G_{\mu\nu}$ is the Einstein tensor.

We consider first the superhorizon case (56) (we generalize the
results of \cite{habaimpd} from $U=\phi$ to general $U$). We
eliminate the environmental fields $\chi_{n}$. Then, the
energy-momentum
\begin{equation}
T_{\mu\nu}^{\phi}=\partial_{\mu}\phi\partial_{\nu}\phi-g_{\mu\nu}
(\frac{1}{2}\partial_{\alpha}\phi\partial^{\alpha}\phi-V)
\end{equation}
is not conserved. We have
\begin{equation}
(T_{\phi}^{\mu\nu})_{;\nu}=\partial^{\mu}\phi(-\gamma^{2}U^{\prime}\partial_{t}U
-\frac{3}{2}\gamma^{2}HU^{\prime}U+a^{-\frac{3}{2}}U^{\prime}\eta)
\end{equation}
The extra terms on the rhs of eq.(85) come from the replacement of
the $\chi $ fields by noise. We can supplement the energy-momentum
 (84) by the dark energy-momentum. So
that the total energy density $\rho_{\phi}+\rho_{de}$ is
conserved. The energy density compensating the non-conservation
law (85) is defined by
\begin{equation}
\partial_{t}\rho_{de}=\gamma^{2}(\partial_{t}U)^{2}
+\frac{3}{4}\gamma^{2}H\partial_{t}U^{2}-a^{-\frac{3}{2}}\partial_{t}U\eta
\end{equation}
Now, if we differentiate the Friedmann equation
\begin{equation}
\frac{3}{8\pi G}H^{2}=\rho_{\phi}+\rho_{de}
\end{equation}
then we obtain the same formula as if there were no $\chi $ fields
(and no noise)
\begin{equation}
\partial_{t}H =-4\pi G(\partial_{t}\phi)^{2}
\end{equation}
Multiplying eq.(56) by $\partial_{t}\phi$ (neglecting the spatial
derivatives) and using eq.(88) we  obtain the conservation law
\begin{equation}\begin{array}{l}
\partial_{t}\Big(-\frac{3}{8\pi G}H^{2}
+\frac{1}{2}(\partial_{t}\phi)^{2}+V-\frac{3}{4}\gamma^{2}HU^{2}
-\frac{\gamma^{2}H}{4\pi
G}(U^{\prime})^{2}\Big)\cr=-\frac{\gamma^{2}H}{4\pi
G}\partial_{t}(U^{\prime})^{2}+\frac{3H\gamma^{2}}{2}\partial_{t}U^{2}
+a^{-\frac{3}{2}}\partial_{t}U\eta
\end{array}
\end{equation}
If we neglect the rhs of eq.(89) then this equation determines $H$
as a function of fields. Eq.(89) gives an alternative formula to
the Friedmann equation (87) for expressing $H$ as a function of
$\phi$ (we can include $U(\phi)$ in the definition of $H$; the
neglect of $(\partial_{t}\phi)^{2} $ in eq.(89), which allows to
determine $H(\phi)$ can be formulated as an additional requirement
for slow roll besides the one discussed below eq.(36)). Assume
that the slow roll conditions as formulated in sec.4 below eq.(36)
are satisfied and eq.(89) defines $H(\phi)$ then we can determine
$a$ as a function of $\phi$
\begin{equation}\begin{array}{l}
a(\phi)=\exp\int^{t}H=\exp(\int^{\phi}d\phi
H(\phi)(\frac{d\phi}{dt})^{-1})=\cr \exp\Big(-\int d\phi H(\phi)
\Big((3H+\gamma^{2}U^{\prime})^{-1}(V^{\prime}+\frac{3}{2}\gamma^{2}HUU^{\prime})\Big)^{-1}\Big)
\end{array}\end{equation}
With the  explicit functions $a(\phi)$ and $H(\phi)$  we have now
a complete probabilistic diffusion model (the noise is determined
by eq.(35)) which allows to determine the Fokker-Planck equation
and calculate the probability distribution of $\phi$ ( for
$U=\phi$ this problem has been investigated  in
\cite{habaimpd}\cite{invmeasure}).

Next we consider a modification of the Einstein-Klein-Gordon
system in the subhorizon case. The formalism of sec.5 is not
useful for an estimate of the dependence of the stochastic
inflanton  equation on the varying $H$ because the approximation
(57) is applicable only for a slowly varying $H$. For estimates of
the subhorizon $k$ we return to eqs.(11)-(13). The solution of the
wave equation for a time dependent $\omega$ is
$\exp(i\int^{t}\omega)$. Then an approximate Green function in
eqs.(11)-(13) is
\begin{equation}
\sin\Big(\int_{s}^{t}\omega\Big)\omega^{-1}
\end{equation}
which for a large $k$ is approximated by \begin{equation}
\sin\Big(k\int_{s}^{t}a^{-1}\Big)k^{-1}
\end{equation}
Inserting in eqs.(2)-(3) the Green function (92) and the wave
functions $\exp(ik\int^{t}a^{-1})$ we obtain in the limit
$\vert\tau-\tau^{\prime}\vert >>\vert {\bf x}-{\bf
x}^{\prime}\vert$ the stochastic equation
\begin{equation}\begin{array}{l}
\partial_{t}^{2}\phi-a^{-2}\triangle \phi+3H\partial_{t}\phi+V^{\prime}(\phi)\cr
+2\pi^{2}\kappa^{2}aU^{\prime} \partial_{t}\int d{\bf y}U(t,{\bf
y})\cr=H^{-1}\Big(\int_{0}^{t}a^{-1}\Big)^{-1}U^{\prime}a^{-1}\eta
\end{array}\end{equation}
where the noise $\eta$ has the covariance
\begin{equation}
\langle \eta(x)\eta(x^{\prime})\rangle=\beta^{-1}\kappa^{2}\pi^{2}
\delta(t-t ^{\prime}) \end{equation} With eq.(93) we can repeat
the argument, which has been based on eqs.(56),(83)-(89), in order
to express $H$ and $a$ in terms of $\phi$. The study of the
Einstein-Klein-Gordon system without the assumption
$\vert\tau-\tau^{\prime}\vert
>>\vert {\bf x}-{\bf x}^{\prime}\vert$ is more involved and
requires a separate investigation because in this case $H$,$a$ and
the noise depend on $\phi$ in a non-local way.

\section{Summary and outlook}
We considered a system of non-linear wave equations describing an
interaction of the inflaton $\phi$ with other scalar fields
$\chi_{n}$. In quantum mechanics if the fields $\chi_{n}$ are not
observable then we average over the states of $\chi_{n}$. As a
result the inflaton is described by a density matrix resulting
from the averaging. The natural state for averaging is the Gibbs
(thermal) state  of maximal entropy for the $\chi_{n}$ fields.
Then, further evolution does not depend on the initial time when
the interaction between the inflaton and the $\chi_{n}$ fields is
switched in. We discussed the model in the classical limit when
the probability distribution is described by the canonical
Hamiltonian of the $\chi_{n}$ fields. In an expanding metric the
canonical Hamiltonian depends on time. However, for large
conformal time in models with an expansion close to the
exponential the time-dependence disappears. The average over the
environmental fields is reduced to an average over the initial
values of the $\chi_{n}$ fields in the Gibbs state. There is some
arbitrariness in the choice of the environment; the number of
fields $\chi_{n}$, the couplings $\lambda_{n}$, the masses $m_{n}$
and the interaction $\lambda_{n}\chi_{n} U(\phi)$ (we choose an
interaction linear in the environmental fields). We discuss
evolution of the inflaton field in an external expanding metric in
two regimes. First, we choose the increasing masses proportional
to couplings and assume negligible (superhorizon)  momenta $k<<aH$
of an infinite set of the environmental fields. These assumptions
after an averaging over $\chi_{n}$ lead directly to a Markovian
wave equation with a friction and white noise. In such a case the
quantum evolution is described by a master equation of the
Lindblad type. Another (subhorizon) limit discussed in this paper
involves large momenta $k>>aH$ and $k>>m_{n}$. The number of
fields $\chi_{n}$ can be finite. After an averaging we obtain a
wave equation for the inflaton which is non-local and
non-Markovian in the friction and noise. If we restrict ourselves
to correlations $\langle\phi(\tau_{1},{\bf
x}_{1})....\phi(\tau_{n},{\bf x}_{n})\rangle$ such that
$\vert\tau_{j}-\tau_{k}\vert >>\vert {\bf x}_{j}-{\bf x}_{k}\vert$
then the stochastic wave equation has  a Markovian limit as a
diffusion wave equation with friction. This equation is different
from the wave equation derived for low momenta of the
environmental fields. Both equations may coincide in a linearized
version after a spatial averaging. In the last section we
discussed a determination of the expansion scale factor $a$ from
the Einstein-Klein-Gordon system of equations. In the superhorizon
case it is possible to obtain a stochastic wave equation with
$a(\phi)$ and $H(\phi)$ dependent on $\phi$ and the noise
independent of $\phi$. In such a case we can calculate all
expectation values on the basis of the Fokker-Planck equation. An
interesting problem for further studies is the dependence of slow
roll conditions and spectral indices on the interaction $U(\phi)$.
In the subhorizon case the noise and $a(\phi)$ depend in a
non-local way on $\phi$. Only in the limit of small spatial
distances ($\vert {\bf x}_{j}-{\bf x}_{k}\vert
<<\vert\tau_{j}-\tau_{k}\vert $) we are able to derive a Markovian
stochastic wave equation with $a(\phi)$ and $H(\phi)$ determined
by Einstein equations.

In this paper we concentrate  our attention on the stochastic
equations satisfied (in various regimes) by the inflaton
interacting with an environment. The environment consists of
superhorizon or subhorizon modes of the classical thermal
environmental fields. In \cite{star}\cite{vilenkin} the subhorizon
modes of the quantum field have been treated as the environment.
The environmental modes are approximated by a noise. If the noise
is approximately Markovian then the quantum states satisfy the
irreversible Lindblad equation. It is well-known that Lindblad
equation leads to decoherence (meaning a destruction of
interference of quantum states).  The interference is a
substantial obstacle to the classical limit. In the standard model
 the large scale structure is formed from quantum
fluctuations \cite{starex}\cite{mukhanov} in the early universe.
Hence, in the meantime quantum fluctuations should become
classical. In \cite{albrecht}\cite{star1} it is shown that if
initially the universe starts from the Gaussian state then with
the time evolution determined by a linear perturbation theory the
resulting squeezed state becomes classical. The measured CMB
fluctuations are indistinguishable from classical fluctuations. At
present, all observations \cite{a1}\cite{a2} show no departure
from Gaussian correlations. However, it may be that prospective
CMB measurements can reveal some non-Gaussian behavior. Such CMB
results could show a destruction of  interference of states in the
early universe and the role of the environment in their
decoherence as discussed in \cite{star2}. The noise in
Einstein-Klein-Gordon system plays at least two roles:it leads to
decoherence and it determines the power spectrum (as calculated in
\cite{starquant} with the quantum noise taking into  account
quantum gravitational fluctuations). The detailed analysis of the
contribution of the noise from the environment to the power
spectrum and to higher order correlation functions of the
inflanton could give important information about the quantum state
in an early universe and its classical limit.

\end{document}